\documentclass{llncs}

\usepackage{graphicx}
\usepackage{amsmath, amssymb}
\usepackage{mathrsfs}
\usepackage{multirow}
\usepackage{enumitem}
\usepackage{pifont}
\usepackage{caption}
\graphicspath{ {figs/} }

\usepackage{url}
\urldef{\mailsa}\path|{swijewickrem@, xingjunm@student., baileyj@,|
\urldef{\mailsb}\path|gek@, sjoleary@}unimelb.edu.au|
\newcommand{\keywords}[1]{\par\addvspace\baselineskip
\noindent\keywordname\enspace\ignorespaces#1}

\begin{document}
\setlength{\textfloatsep}{10pt}
\setlength{\intextsep}{10pt}
\mainmatter  

\title{Feedback Techniques in Computer-Based Simulation Training: A Survey}


\author{Sudanthi Wijewickrema \inst{1}
\and Xingjun Ma \inst{2}
\and James Bailey \inst{2}
\\
 Gregor Kennedy\inst{3}
\and Stephen O'Leary\inst{1}
}


\institute{Department of Otolaryngology\\
\and
Department of Computing and Information Systems \\
\and
Graduate School of Education \\
The University of Melbourne, VIC 3010, Australia\\
\mailsa\\
\mailsb\\
}

\maketitle

\begin{abstract}
Replication or simulation of a real-world scenario for the purpose of training novices in a given profession is not new. For example, the use of cadavers in surgical education has been common practice for centuries. With the advent of computer-based technologies such as virtual reality, simulation has become even more useful as a training platform. Computer-based simulation training (CBST) is gaining popularity in a vast range of applications such as surgery, rehabilitation therapy, military applications, and driver/pilot training, as it offers a low-cost, easily-accessible and effective training environment. Typically, CBST systems comprise of two essential components: 1) a simulation environment that provides an immersive and interactive learning experience, and 2) a feedback intervention system that supports knowledge/skill acquisition and decision making. The simulation environment is created using technologies such as virtual or augmented reality, and this is an area which has gained much interest in recent years. The provision of automated feedback in CBST however, has not been investigated as much, and thus, is the focus of this paper. Feedback is an essential component in learning, and should be provided to the trainee during the training process in order to improve skills, to correct mistakes, and most importantly, to inspire reasoning and critical thinking. In CBST, feedback should be provided in a useful and timely manner, ideally in a way that mimics the advice of an experienced tutor. Here, we explore the provision of feedback in CBST from three perspectives: 1) types of feedback to be provided, 2) presentation modalities of feedback, and 3) methods for feedback extraction/learning. This review is aimed at providing insight into how feedback is extracted, organized, and delivered in current applications, to be used as a guide to the development of future feedback intervention systems in CBST applications.
\keywords{Feedback technique, Computer-based simulation training, Artificial intelligence}
\end{abstract}

\section{Introduction}
Computer-based simulation training (CBST) is a modern training methodology that involves practise and learning in a virtual (as opposed to physical) simulated environment \cite{bell2008current}\cite{lateef2010simulation}. CBST platforms have been able to replicate real-world scenarios in an immersive and interactive fashion \cite{steuer1992defining}, and as such, provide realistic and versatile environments where individuals can learn new information, directly apply this information to simulated tasks, and master complex skills \cite{billings2012efficacy}\cite{menaker2006harnessing}. CBST has many benefits over the traditional way of training: it is low cost/risk, easily-accessible, repeatable, and versatile. Thus, CBST environments are gaining popularity over other simulation methods such as costly and cumbersome physical models. Such applications include surgery \cite{zhao2011can}\cite{furqan2011surface}\cite{masimulation}, driver/pilot training \cite{de2011effect}\cite{howard2011simulation}, military training applications \cite{cosma2011implementing}\cite{howard2011simulation}, and rehabilitation therapy \cite{chen2009psychological} to name a few.

\section{Feedback in CBST}
\label{sec:feedback}
Studies show that supporting the learning process through appropriate and timely feedback is important due to three primary reasons \cite{billings2012efficacy}\cite{davis2005interactive}: 1) feedback can lead to increased motivation by showing discrepancies between current performance and desired level of performance, 2) it can reduce the uncertainty of how an individual is performing, and 3) it can help an individual learn how to correct mistakes. Feedback can be provided in the form of one-to-one instruction or general class room instruction. One-to-one instruction incorporates an adaptive form of feedback that changes dynamically in response to the student's abilities, skills, and other personal differences \cite{billings2012efficacy}\cite{park2003adaptive}. It thus has the flexibility to meet the needs of individual students and has been found to be more beneficial than general class room instruction\cite{bloom19842}. Computer-based training systems implement this method of adaptive feedback by monitoring the performance of the student and delivering suitable feedback using artificial intelligence techniques \cite{park2003adaptive}. 


Feedback intervention in CBST should supplement the sensory information the user gains by interacting with the system and promotes the learning of the intended skill. The sensory information that the user receives from interacting with the virtual environment is \textit{inherent} (intrinsic) feedback and results from the execution of an action \cite{zhou2014real}\cite{schmidt1998motor}. Inherent feedback could be in the form of auditory, visual, or tactile sensations. For example, in a surgical simulator, the user experiences the hardness of the bone and the vibrations of the drill through a haptics device and sees the bone being drilled away and the underlying anatomical structures being uncovered as a consequence of the drilling. Such feedback is usually rich and varied, containing substantial information regarding performance \cite{schmidt1998motor}. In most cases, inherent feedback requires almost no further evaluation, as the user is able to inherently grasp its meaning. However, some aspects of such feedback do not offer immediate recognition of its meaning, but require the user to `learn' or to have some level of prior knowledge. It is thought that inherent feedback is compared to a learned reference of correctness, with this reference acting in conjunction with the feedback in an error-detection process \cite{schmidt1998motor}. In this paper, we consider the provision of inherent feedback to be part of the creation of the simulation environment, and as such, do not investigate it further.

On the other hand, additional feedback that is provided specifically to support the acquisition of desired skills, or \textit{augmented} (extrinsic) feedback is the focus here. This form of feedback intervention should support the processes at the core of learning: selecting (focusing on the relevant information), organizing (forming mental representations of the information), and integrating (combining the new information with previous knowledge existing in long term memory) \cite{billings2012efficacy}\cite{clark2008learning}. Augmented feedback does not assume a reference of correctness and can take the form of an expert trainer. Typically, it is information provided about the action that is supplemental to or augments the inherent feedback \cite{schmidt1998motor}. Such feedback can be in the form of auditory or visual cues that provide the user with an idea of his/her performance. For example, in a surgical simulator, the user can be instructed on different aspects of his/her surgical skill or warned when getting too close to an anatomical structure. The rest of the paper explores in detail different aspects that need to be considered when providing augmented feedback in CBST systems.


\section{Types of Augmented Feedback}
Augmented feedback is a powerful method of conveying information to the user and takes the role of an expert tutor. In the past decades, researchers have examined the various aspects associated with the informational content of the augmented feedback, the schedules to present this information, the temporal placement of this feedback, and plausible theoretical explanations that describe how feedback affects skilled performance \cite{schmidt1998motor}\cite{young2001skill}. In the rest of this section, we investigate augmented feedback as applied to CBST systems with respect to its time of delivery, information content, and presentation methods.

\subsection{Temporal Placement of the Feedback}
There are no universal guidelines for the temporal placement of feedback and it is highly dependent on the type of task and simulation environment involved. However, there are two main categories augmented feedback can be divided into depending on time of delivery: \textit{concurrent} and \textit{terminal}.

Concurrent feedback is delivered as and when an action is being performed, while terminal feedback is delivered at the end of an action \cite{billings2012efficacy}. Note that the terms \textit{formative} and \textit{summative} used in education research refer to concurrent and terminal feedback respectively. As a prerequisite to delivering feedback, the procedure should be divided into tasks or actions. For example, in driver training, we can divide the course into different tasks such as lane-keeping and safe-following-distance maintaining. In the lane-keeping task, we can deliver the distances to left and right lanes as concurrent feedback, while driving skill analysis can only be delivered after a certain period of driving.

Terminal feedback is further divided into \textit{immediate} feedback which is delivered directly after an action has occurred, and \textit{delayed} feedback which is delivered only after a certain interval \cite{schmidt1998motor}. For example, in a surgical simulation that consists of multiple surgical stages, results of a stage should be displayed immediately after the completion of that stage, while skill analysis results may not be displayed until an acceptable level of confidence in its accuracy is reached. The feedback delay could be globally defined (show feedback at a predefined frequency), or it could be with respect to an action (display feedback after a predefined time interval after an action). Ideally, the delay interval could be changed adaptively according to the importance of the feedback, the confidence level of the accuracy of the feedback, and the skill level of the user.

Feedback presented mid-task places more demand on the learner because attention and cognitive resources have to be shifted from the task to interpret the feedback, leading to performance decrements \cite{billings2012efficacy}\cite{munro1985instruction}. In contrast, terminal feedback (especially delayed feedback) mirrors real world scenarios more closely and has been found to be an effective instructional intervention, especially if the task is complex or occurs in real-time \cite{billings2012efficacy}\cite{astwood2008impact}. However, this cannot be generalised, as a hard and fast rule for the temporal placement of feedback as it is dependent on factors such as task, application, and skill level \cite{hattie2007power}\cite{ju2007effectiveness}.

\subsection{Content of the Feedback}
The content being presented and its specificity is again system specific and should ideally be dependent on the skill level of the user as well. Feedback specificity is defined as the level of information presented in the feedback \cite{goodman2004feedback}. Research shows that \textit{detailed} feedback is more useful in learning than \textit{general} feedback that provide broader conceptual suggestions \cite{shute2008focus} \cite{hattie2007power}.

The content of the feedback could also be \textit{descriptive}, providing details of the user's performance or it could be \textit{prescriptive}, with advice on what to do next and how to correct errors that may have occurred. Prescriptive feedback is usually presented when an error has occurred and therefore has negative connotations (see \cite{amorose2003feedback} for an investigation of prescriptive feedback). Although more detailed feedback is generally better, it can introduce a higher level of complexity to the feedback (see \cite{shute2008focus} for details). Therefore, it is essential that a balance be struck between the level of detail and the complexity of the feedback.

Another aspect in the feedback content is the task or tasks it pertains to. Feedback could be \textit{distinct} and represent the performance of each task separately or be \textit{accumulated} that represent a collection of past performances \cite{billings2012efficacy}. For example, in a surgical simulator, once a surgical stage is completed, we can display detailed information about skill analysis, outline errors and suggest ways of overcoming them with respect to the user's accumulated performance in the previous stage. In contrast, we could also deliver skill analysis and feedback after each cut/drill in a briefer and descriptive manner.

The paradigms of \textit{knowledge of results} (KR) and \textit{knowledge of performance} (KP) describe another important dimension of augmented feedback. Knowledge of results feedback provides information about the user's outcome with respect to the environmental goal, while knowledge of performance feedback provides information about the actions of the user with respect to an expert model \cite{billings2012efficacy}. KR feedback informs the user if a goal has been achieved or not, but does not give information about his/her performance, which is the domain of KP feedback.

The feedback being provided could be \textit{quantitative} or \textit{qualitative}. Quantitative (or numeric) feedback while being highly informative, requires a greater level of analysis on the part of the user. Qualitative feedback on the other hand is easier to understand as it is at a higher level and has already been analysed or processed. Studies have found that in the early stages of learning qualitative feedback plays a greater role while in the latter stages quantitative feedback is more important \cite{kilduski2003qualitative}\cite{swinnen1996information}. Therefore, the decision to present quantitative or qualitative feedback should ideally depend on the skill level of the user. If the skill level of the user is detected to be close to that of an expert, quantitative feedback could be displayed while for a novice, qualitative feedback could be provided.

\subsection{Bandwidth Feedback}
Providing qualitative feedback usually requires a level of correctness, deviation from which would elicit a response from the system. This feedback which depends on whether performance is within or outside a preset tolerance limit is called \textit{bandwidth feedback} \cite{de2011effect}. Bandwidth feedback is closely related to the feedback delay, in that it can be used in place of delayed feedback in some situations. For example, in a surgical simulator, instead of presenting skill feedback after a delay, we can display it only if the quality of the surgical technique is found to be outside of the predefined tolerance level.

Bandwidth feedback has been used as an effective learning tool since Thorndike's experiments in 1927 \cite{thorndike1927law}. It could be \textit{on-target} (or \textit{positive}) feedback that the user receives when he/she is doing the `right' thing, or \textit{off-target} (or \textit{negative}) feedback he/she receives if outside of an acceptable performance range. Studies show that on-target feedback reinforces positive habits and could be more rewarding and stimulating for the user while off-target feedback is less likely to distract the user and make him/her dependent on the feedback \cite{de2011effect}.

The choice of on-target or off-target bandwidth feedback (or a combination of the two) depends on the simulation system as well as the complexity and type of the task. For example, in the lane-keeping task in a driving simulator, if the user were provided with on-target feedback when he/she is driving in the correct area, it may cause a sensory overload. Instead, it would be more effective if the user is given feedback only when he/she deviates too much from the lane center.

\subsection{User Interaction in the Feedback}
Allowing user interaction in the feedback system places the decision of what and when feedback should be displayed in the hands of the user \cite{wylie2009active}. From the perspective of the system, this is \textit{passive} feedback where the system only responds when the user requests feedback. In contrast, the system could provide \textit{active} feedback deciding when and what feedback to display. Ideally this decision should be made taking the type of feedback and the level of skill of the user into consideration. For example, in a surgical simulator, an expert user whose goal is to practice surgical procedures would require minimal feedback while a novice user will need step-by-step guidance. In such situations, allowing the user to choose a level of feedback presentation would be beneficial.

Another related aspect of on-demand feedback is the inclusion of \textit{expert stories} in the simulator \cite{guralnick2009putting}. Expert stories are videos that explain a concept, feature, or procedure and the user should have access to them at various important points of their simulated experience (examples can be found in \cite{guralnick2009putting}). 

\subsection{Presentation of the Feedback}
The way in which feedback is presented also plays a major role in a CBST system. The method of presentation should depend on the system, the task being performed, and the nature and content of the feedback being presented. In this context, different aspects such as the method of presentation and the spacial placement of the feedback should be considered.

Depending on the type of feedback, the decision needs to be made as to which sensory input (\textit{visual} or \textit{auditory}) should be used to present the feedback. Most complex tasks are mastered in response to information perceived visually \cite{sigrist2011visual}. However, visual feedback may overload the capacities of visual perception and cognitive processing (for a detailed investigation on how visual and auditory concurrent feedback affected the learning of complex motor tasks, see \cite{sigrist2011visual}). Therefore, care needs to be taken in deciding which form of presentation is used for which type of feedback. For example, a simple beep could indicate that an action may cause potential danger, while visual feedback can be used to show the consequences when a destructive action has been performed.

\textit{Verbal} feedback is presented in a form that is spoken or is capable of being spoken \cite{schmidt1998motor} (could be auditory or visual). In contrast, \textit{non-verbal} feedback is `symbolic' and may require less attention for the user to understand. For example, we could indicate the closeness to an object by the use of a `light' that is usually green but turns red when the user gets too close.

Studies show that the location of the information serves as an important aspect of spatial memory \cite{rothkopf1971incidental}\cite{tan2001infocockpit}. As such, it plays a major role in the learning of skills and therefore, in the presentation of feedback in a CBST system as well. 

\textit{Head-down} presentation is when the feedback is displayed outside of the user's field of view and therefore requires his/her attention to be split between the task and the feedback. In contrast, \textit{head-up} displays are in the field of view of the user and therefore overcome the above issue. However, they introduce `clutter' to the scene and this can affect the performance of the user adversely. Studies have been conducted as to the effectiveness of the two types of displays with respect to different applications (for example, driving performance \cite{liu2004comparison}).

\textit{Peripheral} displays on the other hand remain at the edge of the user's main focus of attention and allow him/her to be aware of information without being overburdened \cite{matthews2007defining}\cite{weiser1997coming}. Hence, peripheral displays can be defined in terms of their spacial placement (in the periphery of the user's field of view) and their purpose (to be unobtrusive when providing information). \textit{Interruptive} displays in contrast disrupt the user from his/her primary action, and cause concentration to be switched to the information being presented \cite{matthews2007defining}.

From another perspective, displays can be categorized into \textit{ambient} and \textit{notification} displays. Ambient displays are typically defined as aesthetic displays, often integrated into the environment, and designed to convey information subtly \cite{matthews2007defining}\cite{stasko2004personalized}. Notification displays, on the other hand, attempt to deliver current, important information to users in an efficient and effective manner without causing unwanted distraction to ongoing tasks \cite{matthews2007defining}.

\section{Feedback Extraction}
In the previous sections, we reviewed various kinds of augmented feedback and presentation methods. In this section, we continue to investigate the techniques that are used to extract augmented feedback. Feedback extraction is to extract the information/instructions needed to be delivered to the user based on existing action samples or predefined rules. In general, feedback extraction methods can be categorized into three types: 1) predefined feedback, 2) rule-based feedback extraction, and 3) automated feedback extraction.

\subsection{Predefined Feedback}
Feedback can be predefined manually based on the learning task and the steps that are required to complete a task. Such feedback can be predefined by a human expert in the field and are often case-specific. For example, in Wijewickrema et al. \cite{wijewickrema2016provision}, step-by-step manually defined instructions were used to train novice surgeons. Each step of the procedure was shown only once the previous step was completed to guide the trainee through a surgery. Another example is the `follow-me' approach used in a dental surgical simulator \cite{rhienmora2011intelligent} . It utilized a carefully selected expert procedure that was displayed as a `ghost drill' that the trainee should follow to learn a procedure. Predefined feedback is an effective way of training a procedure, but is typically only applicable for clearly defined tasks. Also, the feedback is not interactive, and as such, it does not adjust in real-time if the user deviates from the suggested action. 

\subsection{Rule-based Feedback Extraction}
Rule-based (or semi-automated) extraction methods work with a series of predefined rules, the violation or fulfillment of which is used to indicate the level of performance. Relevant feedback is provided depending on the detected skill level. The rules can be constraints that regulate user actions. For example, in the lane-keeping task, when the car deviates too far from the lane center, `back-on-track' feedback will be generated \cite{de2011effect}. In this case, the distance to lane center is a constraint. The rules can also be some standardized expert examples. Rule-based methods have been widely adopted for extracting KR (i.e., \textit{knowledge of results}) feedback. This is because result assessment often involves well-defined rules (or objectives) \cite{blevins2005quantifying}\cite{sewell2005achieving}. The percentage of on-target driving, reaction time and steering reversal rate in a lane-keeping driving simulation \cite{de2011effect} is an example of this.

\subsection{Automated Feedback Extraction}
Both predefined and rule-based methods are task-specific and cannot be easily generalised to other tasks. Also, they require substantial involvements of human experts. This has been one of the obstacles for the spread of CBST as an effective training platform. Recent works attempt to automate the feedback extraction process using data mining techniques so as to increase its generalisability and reduce the reliance on human experts. Data mining techniques are used in a supervised way to train a performance classifier (or discriminator) to separate expert behavior from novice behavior, and extract actionable knowledge from the classification model as feedback to improve novice behavior.

One such example is the use of Dynamic Time Warping (DTW) to classify the time series of surgical actions and support feedback provision in a simulated lumbar disk herniation surgery \cite{forestier2012classification}. The difficulty for time-series based feedback extraction is choosing an appropriate time window that can be used to define the time series of one or multiple sequential actions. Also, it is hard to distinguish experts from novices at the beginning of the task where not much data is available. In a temporal bone surgery simulator, a supervised pattern mining algorithm was used to identify expert/novice behavioural patterns from existing action samples \cite{zhou2013pattern}. With these patterns, the closest expert pattern was delivered to the user as feedback when a novice pattern was detected during training. The challenge for pattern-based methods is that experts and novices often share a considerable amount of similar behavior.

Zhou, et al. investigated the use of a random forest model to extract detailed skill feedback in a surgical simulator \cite{zhou2013constructive}. Similar techniques have been used in customer relationship management as well, to change disloyal customers to loyal ones \cite{yang2003postprocessing}\cite{yang2007extracting}. However, it has been proven that extracting optimal feedback from additive tree models such as random forest and gradient boosted trees is NP-hard \cite{cui2015optimal}. Cui, et al. proposed an approach to transform this NP-hard problem to an integer linear programming (ILP) problem so that it can be solved by ILP solvers \cite{cui2015optimal}. As such, a random forest with thousands of trees will be transformed to a complex ILP problem that has too many constraints. Consequently, this approach is very time-consuming when dealing with large scale random forests and may not be the best option for real-time feedback.

The essence of automated feedback extraction is extracting actionable knowledge from a pre-trained classification model with which a skill instance can be moved from an undesired class to a desired one. Feedback extraction is superior to the classification task in that it not only tells you which class an instance belongs to but also tell you what needs to be done to move an instance to a different class. Therefore, model explanation or interpretation is important in the feedback extraction process. Existing works of this nature include extracting rules from a trained neural network \cite{andrews1995survey}, explaining the predictions of a classifier \cite{ribeiro2016should}\cite{lipton2016mythos}, and exploring the intriguing properties of deep neural networks \cite{craven1996extracting}\cite{szegedy2013intriguing}, to name a few. The adversarial property of neural networks has been explored recently to extract feedback for simulation-based learning \cite{ma2017extracting}. This property indicates that, in deep neural networks, a small amount of feature changes can move an instance to a different class group with high confidence \cite{nguyen2015deep}. Moreover, this property also exists in classification models that are linear in nature \cite{goodfellow2014explaining}. This implies that many classifiers can be used to extract feedback using adversarial techniques.

Most of existing works in this field attempt to solve the feedback extraction problem in a non-cognitive way. In general AI, more and more cognitive models are proposed to solve learning and planning problems such as autonomous driving cars \cite{chen2015deepdriving} and playing the game of Go \cite{silver2016mastering}. These methods directly take the captured images of real-world environments as input and output meaningful instructions. Such models can be applied to CBST to extract feedback. A possible challenge of applying cognitive models to CBST is that the collection of simulation data can be extremely time-consuming. Generative adversarial networks (GANs) can be used to generate expert-like actions or scenes as a supplement to real-world data \cite{goodfellow2014generative}.

\section{Discussion}
Feedback is an important aspect in the learning process. As such, if computer-based simulation training environments are to be used as autonomous training platforms, it is imperative that they come equipped with automated feedback intervention systems. However, this area of research is only now attracting attention, and there is still a lack of automated feedback systems and evidence of their effectiveness in improving skill acquisition. In an effort to guide the development of future feedback intervention systems in computer-based simulation training platforms, this paper reviewed the different aspects of feedback and their presentation modalities, along with methods of feedback extraction. It is expected that with the use of artificial intelligence techniques in the development of automated feedback systems, the usefulness of computer-based simulation training environments will improve, leading to more wide-spread use.

\bibliographystyle{splncs03unsrt}
\bibliography{AI-2017}

\end{document}